\documentclass[aps, prl,superscriptaddress,showpacs,twocolumn,10pt]{revtex4-1}
\pdfoutput=1

\usepackage{amsfonts,amssymb,amsmath}
\usepackage{graphics,graphicx,epsfig}
\usepackage{amsthm}
\usepackage{epstopdf}
\usepackage{placeins}

\renewcommand{\eqref}[1]{Eq.~(\ref{#1})}
\newcommand{\fref}[1]{Fig.~\ref{#1}}

\begin{document}

\title{Experimental observation of current reversal in a rocking Brownian motor}

\author{Christian~Schwemmer}
\affiliation{IBM Research GmbH, S\"aumerstrasse 4, 8803 R\"uschlikon, Switzerland}

\author{Stefan~Fringes}
\affiliation{IBM Research GmbH, S\"aumerstrasse 4, 8803 R\"uschlikon, Switzerland}
\affiliation{Department of Physics, University of Zurich, Winterthurerstrasse 190, 8057 Zurich, Switzerland.}

\author{Urs~Duerig}
\affiliation{IBM Research GmbH, S\"aumerstrasse 4, 8803 R\"uschlikon, Switzerland}
\affiliation{SwissLitho AG, Technoparkstrasse 1, 8005 Z\"urich, Switzerland}

\author{Yu~Kyoung~Ryu}
\affiliation{IBM Research GmbH, S\"aumerstrasse 4, 8803 R\"uschlikon, Switzerland}

\author{Armin~W.~Knoll}
\affiliation{IBM Research GmbH, S\"aumerstrasse 4, 8803 R\"uschlikon, Switzerland}

%\begin{abstract}
%A reversal of the particle current in rocking Brownian motors was predicted more than 20 years ago; however, an experimental verification and a deeper insight into the underlying mechanisms remained elusive. Here, we investigate the high frequency behavior of a rocking Brownian motor for charged nanoparticles based on electrostatic interactions in a 3D shaped nanofluidic slit and electro-osmotic forcing of the particles. A sub ms temporal and $\approx\,10\,$nm spatial resolution of the {60\,nm} gold spheres allows us to measure the time-resolved and frequency dependent evolution of the particle probability density {\it in-situ}. At 250\,Hz the particle current changes sign, in agreement with a theoretical model based on the time-dependent Fokker-Planck equation. From this fit-parameter free description and its excellent agreement with the observed behavior, we trace the origin of the current reversal to the asymmetric and increasingly static probability density at high frequencies.  
%In forward direction the asymmetry leads to a significant delay in the current buildup, resulting in an over-proportional decrease of the forward current followed by current reversal.   
%\end{abstract}

\begin{abstract}
	A reversal of the particle current in overdamped rocking Brownian motors was predicted more than 20 years ago; however, an experimental verification and a deeper insight into this noise driven mechanism remained elusive. 		
	Here, we investigate the high frequency behavior of a rocking Brownian motor for {60\,nm} gold spheres based on electrostatic interaction in a 3D shaped nanofluidic slit and electro-osmotic forcing of the particles.
	We measure the particle probability density {\it in-situ} with {10\,nm} spatial and {250\,$\mu$s} temporal resolution and compare it with theory.
	At a driving frequency of {250\,Hz}, we observe a current reversal which can be traced to the asymmetric and increasingly static probability density at high frequencies.
\end{abstract}


\date{\today}
\maketitle

\emph{Introduction.}---Nature uses fascinating machines, called molecular motors, to achieve intracellular directed transport or to propel bacteria in highly diffusive environments. 
Such devices, that transform random Brownian fluctuations into directed net motion are called Brownian motors. 
Both natural and artificial motors require an asymmetric, ratchet-shaped potential and an unbiased external driving force to bring the system out of equilibrium~\cite{Haenggi2009}.
Depending on whether the transport mechanism is based on a fluctuating potential or on a fluctuating force, the motors are called flashing or rocking Brownian motors~\cite{Astumian1997}, respectively. Over the years, predominantly flashing Brownian motors were realized for various different particle 
systems~\cite{Faucheux1995a, Faucheux1995b, Lee2005, Rosselet1994, Gorre-Talini1997, Marquet2002, Faucheux1995b, Bogunovic2012, Lee2005, Arzola2011}. Only recently, a rocking Brownian motor for nanoparticle transport and separation was implemented, which exploits the particle-wall interaction in a 3D shaped nanofluidic slit to create a static energy landscape and is driven using electro-osmotic forces ~\cite{Skaug2017}.
\newline
\indent
Rocking Brownian motors are highly non-linear devices with an average particle current that depends critically on the driving conditions, a promising feature for nanoparticle separation~\cite{Reimann2002, Bartussek1994,Skaug2017}. 
Perhaps the most striking manifestation of this non-linear behavior is a reversal of the current direction at higher rocking frequencies as theoretically predicted~\cite{Bartussek1994}. Current reversals were observed in several ratchet based systems, such as, e.g. optically trapped micro-particles~\cite{Lee2005,Arzola2011}, cold atoms~\cite{Jones2004} or superconducting circuits~\cite{Beck2005,Shalom2005,Silva2007}. The origin of 
current reversal in these systems was identified to deterministic (inertia) effects  \cite{Lee2005,Arzola2011,Jones2004,Beck2005}, interactions between several particle types~ \cite{Shalom2005,Silva2007}, transitions in transport paths~ \cite{Linke1999}, or interference effects~\cite{Linke1998}. For overdamped rocked ratchets, however, deterministic effects do not apply \cite{Bartussek1994} and particle-particle interactions and more complex effects can be safely ruled out in the highly diluted particle suspension studied here. Thus, for overdamped rocked ratchets current reversal has not been experimentally observed nor systematically explained in spite of their nano-technological potential. 
%However, a particle transport along the `hard' direction of a rocking Brownian motor has not been observed yet.
%However, particle transport of a rocking Brownian motor along the steep slope of its ratchet-shaped potential has not been observed yet.
%Moreover, this current reversal could not be rationalized theoretically by a deterministic particle motion and was vaguely explained as an 'co-operative interplay between noise and deterministic, finite-frequency driving' \cite{Bartussek1994}. 
\newline
\indent
Here, we experimentally observe the {\it time-resolved} evolution of the probability density as a function of the driving frequency in a nanofluidic rocking Brownian motor ~\cite{Skaug2017}. We compare the experimental results with a theoretical model based on a numerical solution of the time dependent Fokker-Planck equation with no fit parameters. We first analyze the behavior at low frequencies, where we discover a finite response time $\tau_R$ of the system to reach the steady state in forward or backward direction. At frequencies similar or higher than $\tau_R^{-1}$ we observe a more rapid decrease of the forward current compared to the backward current, which results in an overall current reversal at $\approx 250\,$Hz. This asymmetric behavior of the current can finally be assigned to the evolving asymmetry of the particle probability distribution, approaching the static profile.

\emph{Rocking ratchets.}---The dynamics of an over-damped Brownian particle in a ratchet potential $V(x)$ with oscillating
external rocking force $F(t)$ is described by a Langevin equation \cite{Langevin1908,Risken1989} of the form
\begin{equation}
\gamma\dot{x} = -\partial_x V(x) + F(t) + \xi(t)
\label{eq:langevin}
\end{equation}
where $\xi(t)$ is a random force obeying the fluctuation-dissipation relation $\langle \xi(t) \xi(s)\rangle = 2 \gamma k_B T \delta(t-s)$. 
For Stokes$'$ drag, the drag constant is given by $\gamma = 6 \pi \eta R$ with dynamic viscosity $\eta$ and particle radius $R$. 
It is well known that the stochastic process associated with \eqref{eq:langevin} can also be expressed 
in terms of a probability density $\rho(x,t)$ satisfying the Fokker-Planck equation
\begin{equation}
\partial_t \rho(x,t) = \partial_x \left[ \left( \frac{1}{\gamma}\partial_x \tilde{V}(x,t) + D_0 \partial_x \right) \rho(x,t) \right]
\label{eq:smoluchowski}
\end{equation}
with $\tilde{V}(x,t) = V(x) - xF(t)$ and diffusion coefficient $D_0=k_{\rm B}T/\gamma$. Note that \eqref{eq:smoluchowski} can also be interpreted 
as a continuity equation
\begin{equation}
\partial_t \rho(x,t) + \partial_x S(x,t) = 0
\label{eq:continuity}
\end{equation}
with probability current $S(x,t) = -( \frac{1}{\gamma} \partial_x \tilde{V}(x,t) + D_0 \partial_x ) \rho(x,t)$.
We focus our attention here on the Fokker-Planck equation because it directly describes the experimentally accessible probability density $\rho(x,t)$.

Except for a few special cases, like e.g. an external force varying only 
slowly with time \cite{Magnasco1993, Risken1989}, \eqref{eq:smoluchowski} cannot be solved analytically. 
For a detailed derivation of a numerical solution see SM1 of the Supplemental Material 
(SM)~\cite{SM}. Briefly, because the potential $V(x)$ is periodic in space with period $L$ and the driving force is periodic in time with angular 
frequency $\omega$,  we chose the ansatz~\cite{Denisov2009}
\begin{eqnarray}
\rho(x,t) &=& \sum\limits_{r,s} u_{r,s}e^{i 2\pi r x/L}e^{i s \omega t}\label{eq:rho}\\
\tilde{V}(x,t) &=& \sum\limits_r b_r e^{i 2\pi r x/L} - x \sum\limits_s a_s e^{i s \omega t}
\label{eq:V}
\end{eqnarray}
with Fourier coefficients $u_{r,s}$, $b_r$ and $a_s$. For a driven system, the total force $-\partial_x \tilde{V}(x,t)$ is both periodic 
in space and time, and $\partial_x \tilde{V}(x,t)$ can thus be expressed as
\begin{equation}
\partial_x \tilde{V}(x,t) = \sum\limits_{r,s} g_{r,s}e^{i 2\pi r x/L}e^{i s \omega t}
\label{eq:Vprime}
\end{equation}
with $g_{r,s} = \frac{i 2 \pi b_r r}{L} \delta_{s,0} + a_s \delta_{r,0}$ \cite{SM}.
Insertion of Eqs.~(\ref{eq:rho}) and (\ref{eq:Vprime}) into \eqref{eq:smoluchowski} delivers, after simplification, the recurrence 
relation
\begin{equation}
\left(i\gamma \omega s + \frac{4\pi^2 D_0 \gamma}{L^2} r^2 \right) u_{r,s} - \frac{i 2 \pi}{L} \sum\limits_{m,n} r g_{m,n} u_{r-m,s-n} = 0.
\label{eq:recurrence}
\end{equation}
The unknown coefficients $u_{r,s}$ can be determined by transforming \eqref{eq:recurrence}
into a matrix equation which can then be solved by standard linear algebra routines~\cite{SM}.

\begin{figure}[t!]
	\centering
  \includegraphics[width=0.48\textwidth]{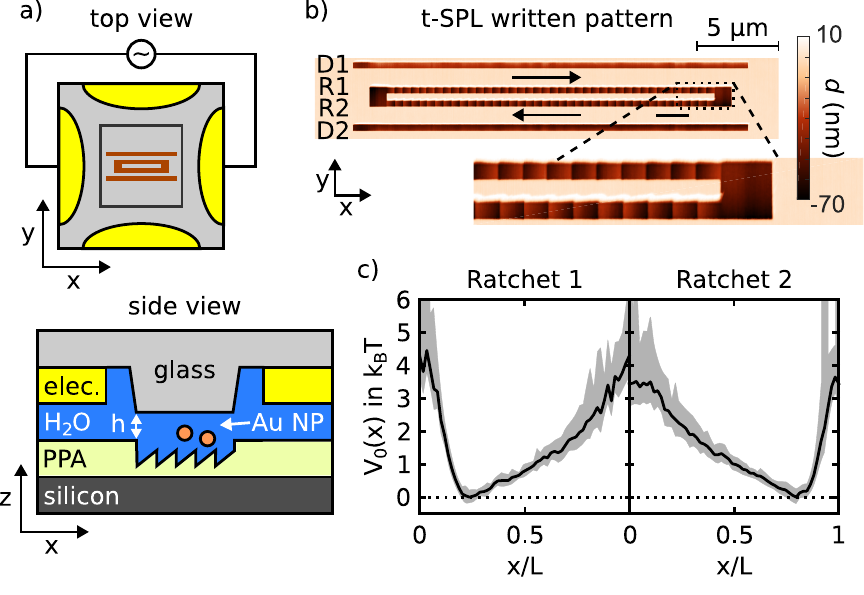}[t!]
	\caption{(color online) a) Schematic top and side view of the nanofludic slit (not to scale). The cover 
	glass exhibits a 30 to 50\,$\mu$m central mesa which allows for optical access and to reduce the gap distance $h$
	down to approximately 100\,nm. To drive the motor, four electrodes were placed around the mesa across which a zero-mean square 
	shaped voltage can be applied. b)~Scanning probe topography image of the patterned ratchets (R1 and R2) in the middle 
	and the drift fields (D1 and D2) next to them. c)~Experimentally determined ratchet potentials for ratchets R1 and R2 (straight lines)
	together with one standard deviation (shaded area).}
	\label{fig:fig1}
\end{figure}

\emph{Experimental implementation.}---The experimental apparatus described in \cite{Fringes2016,Skaug2017} was used to confine 60\,nm gold 
nanoparticles in electrolyte to a nanofluidic slit of controllable size (Debye length $\kappa^{-1}$ $\approx$ 10\,nm, see SM2~\cite{SM} for more details on the nanoparticles). 
The confining surfaces consisted of glass and the thermally sensitive polymer polyphthalaldehyde (PPA), see \fref{fig:fig1}a) and SM3~\cite{SM}. 
Thermal scanning probe lithography~\cite{Garcia2014} was used to fabricate two ratchets, R1 and R2, in PPA with a sawtooth profile of 30 to 60\,nm depth and a period of $\approx$\,$600$\,nm forming a continuous racetrack, see \fref{fig:fig1}b).
As both confining surfaces and the nanoparticles themselves were negatively charged, the patterned topography directly translated into an electrostatic energy landscape with a modulation of several $k_{\rm B}T$ for slit sizes of a few Debye lengths $\kappa^{-1}$ 
\cite{Krishnan2010,Kim2014,Skaug2017}.

\begin{figure}[t!]
	\centering
	\includegraphics[width=0.48\textwidth]{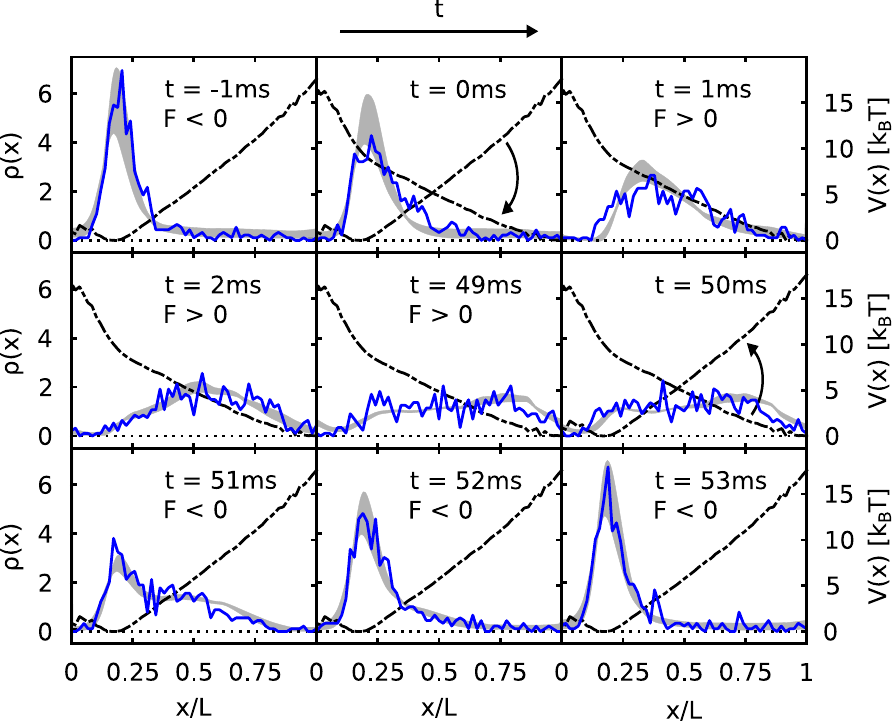}[t!]
	\caption{(color online) Time evolution of the probability density (blue) inside ratchet R1 for square wave driving at 10\,Hz and 4\,V. 
		At $t=0$\,ms, the force changes from negative to positive and at $t=50$\,ms it changes back to negative again. The two
		configurations correspond to a forward and backward tilt of the potential (dashed dotted line). After each sign change, 
		the motor needs $\approx$\,$4$\,ms to follow the change in force. The gray shaded areas show the theoretically expected evolution 
		taking into account the uncertainties in determining the tooth height, the diffusivity and the rocking force.}
	\label{fig:fig2}
\end{figure}

\begin{figure}[t!]
	\centering
	\includegraphics[width=0.48\textwidth]{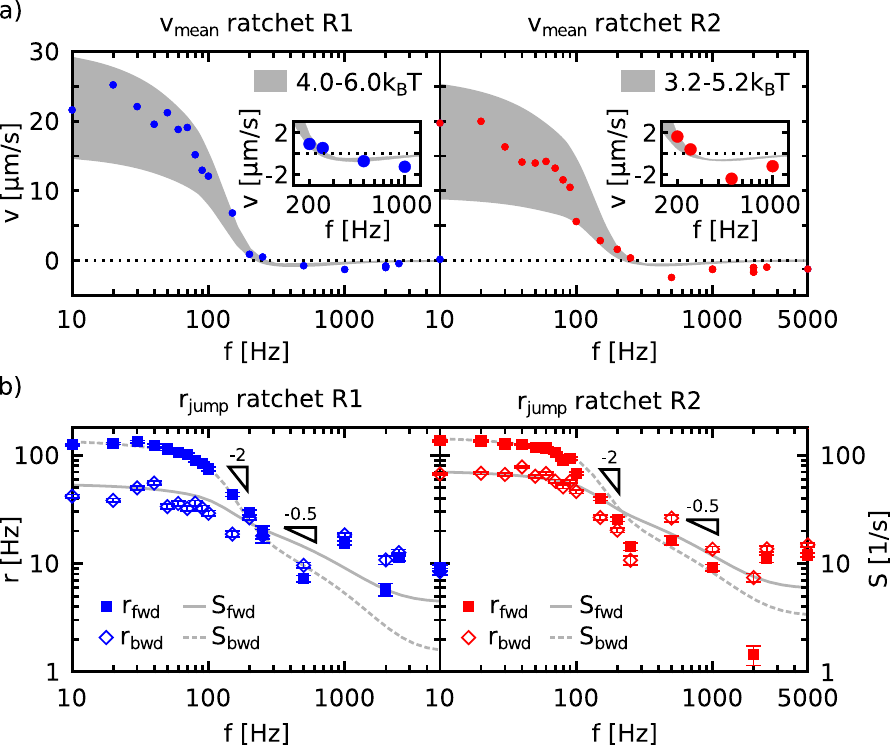}
	\caption{(color online) a) Average particle drift speeds in ratchet R1 (blue) and R2 (red) for different driving frequencies 
		in comparison with the theoretically expected speeds (gray shaded area). At a frequency of $250\,$Hz, the drift
		direction reverses sign. The error bars are smaller than the marker size.
		Inset: Close-up of the frequency range where the current reversal occurs. b) Forward and backward jump rates
		as determined from the particle trajectories in comparison with the expected rates calculated for tooth heights of 5.5\,$k_{\rm B}T$ and 5.0\,$k_{\rm B}T$. 
		Note that both tooth heights are in good agreement with the potentials shown in \fref{fig:fig1} when the
	  	measurement inaccuracy is taken into account. The error bars of the experimental jump rates represent the 
Poissonian 
	  	error of the number of jumps.  }
	\label{fig:fig3}
\end{figure}

In a first step, the gold nanoparticles were loaded to the apparatus, the gap distance was reduced to $h=95.6 \pm 1.5$\,nm and two nanoparticles 
were trapped in ratchet R1 and three in ratchet R2 (see SM4 and SM5 for details on setup stability and particle numbers~\cite{SM}).
Using interferometric scattering detection (iSCAT)\cite{Lindfors2004, 
Fringes2016}, the particle positions were recorded with a high frame rate camera (MV-D1024–160-CL-12, Photon Focus) allowing 
for a frame rate of 4000\,fps and an effective illumination time of $\approx$\,$30$\,$\mu$s \cite{Fringes2018}. Particle detection and tracking 
were carried out using Trackpy for Python \cite{Trackpy} which builds on the Crocker-Grier algorithm \cite{Crocker1996}.
Centre positions of the particles were assigned to the centre of the recorded diffraction limited particle images. The short illumination time results in a detection accuracy of $\lessapprox 10$\,nm \cite{Fringes2018}.
The obtained 2D particle positions inside the ratchets were first collapsed on the x-axis and then, by overlaying the single teeth, a 1D histogram of particle positions for an average ratchet tooth was determined. This normalized histogram can be interpreted as the experimental probability density $\rho(x)$ to find a particle at position $x$ within a ratchet tooth~\cite{Skaug2017} (see also SM6~\cite{SM}). In the static, non-driven case, the interaction potential of a single tooth was obtained by applying Boltzmann's principle $\rho(x) \propto e^{-V(x)/k_{\rm B}T}$, see \fref{fig:fig1}c). For ratchets R1 and R2, we measured an average tooth height of $4.5\,k_{\rm B}T$ and $3.7\,k_{\rm B}T$, respectively. Although more than 50000 particle positions where recorded, the standard deviation is still $> 1\,k_{\rm B}T$ at the potential maxima, see \fref{fig:fig1}c), due to $\rho(x) < 0.1$ and thus a low probability to find a particle at these positions.
\newline
\indent
To cross-check our results, we measured the particle jump rate $r$ between neighboring teeth and compared it against the expected 
rate as a function of tooth height according to Kramer's escape theory. The experimental jump rate was $5.5 \pm 0.7$\,Hz for ratchet 
R1 (see also SM7~\cite{SM}). In the over-damped limit, the jump rate in a periodic potential
is directly linked to the effective diffusion coefficient by $r = 2 D_{eff}/L^2$~\cite{Ferrando1992}. As shown by Lifson and 
Jackson~\cite{Lifson1962}, $D_{eff}$ can be expressed in terms of the free diffusion coefficient $D_0$ 
and $V(x)$ as $D_{eff} = D_0 / \left[ \int_0^L \exp\left(\frac{V(x)}{k_{\rm B}T}\right) \int_0^L \exp\left(-\frac{V(x)}{k_{\rm B}T}\right) \right]$.
With an experimental diffusion coefficient $D_0 = 4.3 \pm 0.2 \mu$m$^2/$s, the expected jump rates for rescaled teeth could easily be 
calculated. The best agreement for ratchet R1 was achieved for a tooth height between $3.9\,k_{\rm B}T$ and $4.4\,k_{\rm B}T$. 
Note, that we assume $D_0$ to be constant in this calculation. However, due to the gap distance variation between 130 and 160\,nm in the ratchet, we expect $D_0$ to decrease by $\approx 10\%$ in the most confined space \cite{Fringes2018} (see SM8~\cite{SM} for a detailed discussion). This effect may account for the slightly lower values of the calculated potential.
For ratchet R2, the jump rate could not be determined reliably as two particles were trapped in close proximity, often visiting the same cell, where they cannot be distinguished in the diffraction limited optical setup. This led to many short trajectories and an unreliable determination of the rate (see  SM7~\cite{SM} for further details).
%The two particles often came close together and then only a single image was observed which impeded single particle detection. Accordingly, it was not possible to distinguish between two particles that passed by each other and two particles that were repelled from each other (see  SM7~\cite{SM} for further details).
Nonetheless, there is good agreement between the two approaches for ratchet R1 and the experimental potentials of ratchet R1 and R2 shown in \fref{fig:fig1}c) cannot be distinguished within the limits of their errors.
Therefore, for the rest of our discussion, we used their rescaled average for modeling (see SM6~\cite{SM}).
\newline
\indent
To power the motor, we created non-equilibrium fluctuations by applying a square wave voltage across the 
electrodes, see \fref{fig:fig1}a). The force on the particles results predominantly from the electro-osmotic plug flow in the slit \cite{Skaug2017}.
To quantify the strength of the force $F$ we measured the average speed $\langle v_{drift} \rangle$ 
of the particles in the drift fields D1 and D2. By combining Einstein's relation $D_0=k_{\rm B}T/(6\pi \eta R)$ with Stoke's equation for particle drag $F = 6\pi \eta R \langle v_{drift} \rangle$, the force is given by $F =  k_{\rm B}T \langle v_{drift} \rangle/D_0$. For our fixed amplitude of 4\,V we measured a force of $F=28.9 \pm 0.7 k_{\rm B}T/\mu$m ($109 \pm 3\,$fN), independent of the driving frequency. Notably, since the force is exerted by the laminar plug flow, we also expect the force to be $\approx 10\%$ larger in the most confined space, which cancels the effect of the reduced $D_0$ on  $v_{drift}$.
\newline
\indent
Owing to the high temporal resolution, we can directly investigate the response of $\rho(x,t)$ upon a sign change in $F$ in 1\,ms steps. For ratchet R1 and a frequency of $f=10$\,Hz, $\rho(x,t)$ (blue) and $\tilde{V}(x,t)$ (dash-dotted line) are shown and compared to the theoretical results (gray area) in \fref{fig:fig2} (see S1, S2 and SM9 of the SM~\cite{SM} for a slow motion movie of the experiment, the time evolution of $\rho(x)$, and for ratchet R2). Initially, $F$ is negative
and the ratchet potential is tilted towards the steep slope (backwards direction) exhibiting a local minimum with a barrier of $1.6$\,$k_{\rm B}T$ to the left. The particles are localized at the minimum leading to a pronounced maximum of $\rho(x,t)$, 
see top left panel of \fref{fig:fig2}. At $t=0$\,ms, the force changes sign and the ratchet potential is tilted towards 
the shallow slope (forward direction). As a result,  
the peak in $\rho(x,t)$ drifts in positive $x$ direction and broadens due to diffusion. For both effects we calculate the characteristic time scales. Drift along an average potential slope of $16 k_{\rm B}T/0.59$\,$\mu$m~$\approx$~$27 k_{\rm B}T/$\,$\mu$m requires a time of $t_D = L  k_{\rm B}T/F D_0 = 5$\,ms to cover a distance of $L$. The roughly Gaussian peak in $\rho(x,t)$ of amplitude $A_0$ evolves into one with amplitude $A_1$ in $t_B = (A_0^2 - A_1^2)/(4\pi D_0 A_0^2 A_1^2)$~\cite{Risken1989}. For $A_0 = 7$ and $A_1 = 2$, one obtains $t_B \approx 4$\,ms. Both values agree with the observed system relaxation time to reach a steady state of $\tau_R \approx 4\,$ms, see also the buildup of the peak in backward direction within $3-4\,$ms. 
\newline
\indent
We expect nonlinear behavior when the half-period $T_f/2$ of the driving frequency is of similar duration as $\tau_R$. Indeed, we find that the average particle drift $v$ starts to decrease sharply at  $f \approx 100\,$Hz from its low-frequency value of $\approx$\,20\,$\mu$m/s, see \fref{fig:fig3}a). Moreover, at 250\,Hz we observe that the current changes sign and reaches values of -1 to -3\,$\mu$m/s at 500\,Hz.
The same behavior is reflected in the forward and backward cell to cell jump rates, which were obtained from the particle trajectories, see \fref{fig:fig3}b). For slow rocking, the backward jump rate $r_{\rm bwd}$ and the forward jump rate $r_{\rm fwd}$ are almost constant and $r_{\rm fwd}$ exceeds $r_{\rm bwd}$ by $\approx$\,70\,Hz. For $f > 100$\,Hz however, the two rates show a remarkably different scaling behavior. $r_{\rm fwd}$ decreases  $\propto$\,$f^{-2}$ whereas $r_{\rm bwd}$ decreases $\propto$\,$f^{-0.5}$, resulting in the current reversal at 250 Hz. 

\begin{figure}[t!]
	\centering
	\includegraphics[width=0.48\textwidth]{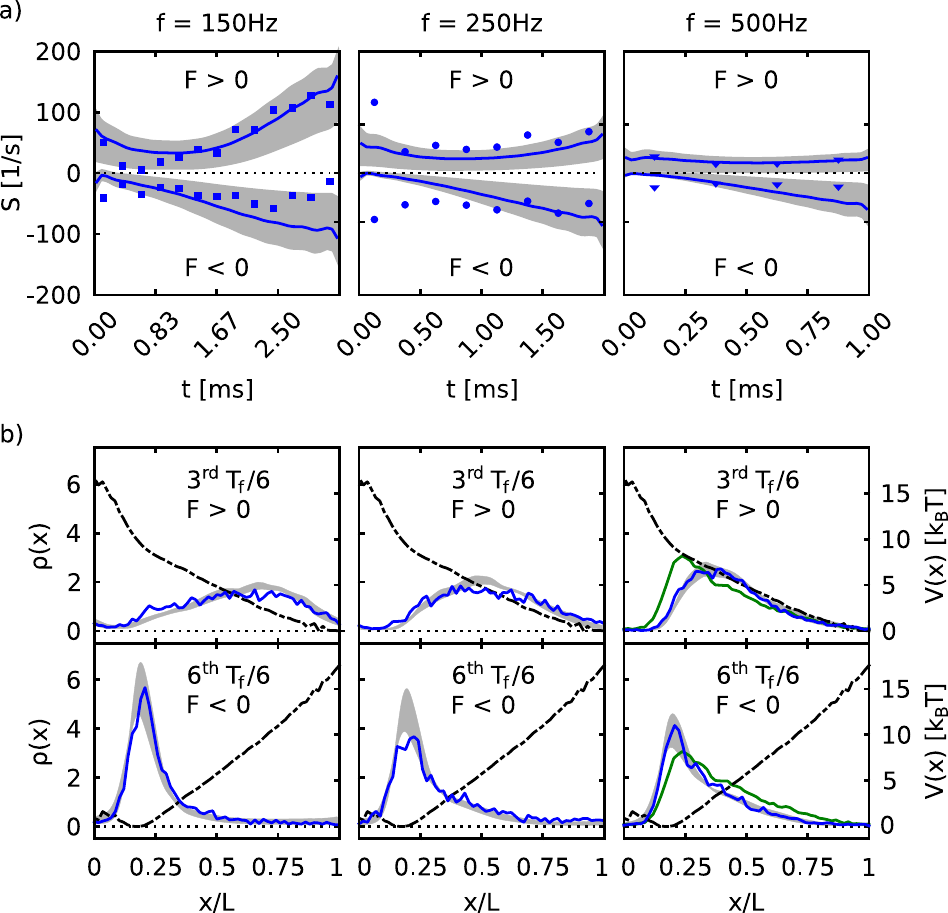}[t!]
	\caption{(color online)  a) Buildup of the current $S$ in ratchet R1 at the maximum of $V(x)$ for frequencies of 150\,Hz (square), 
		250\,Hz (circles) and 500\,Hz (triangles) in comparison with our theoretical model (blue lines) taking into account
		experimental uncertainties (gray shaded area). When $F>0$, the potential is tilted towards the shallow slope whereas for $F<0$, the potential is tilted towards the steep slope.
		%The backward current outgrows the forward current above 250\,Hz leading to a current reversal.
		b) Probability densities averaged over $T_f/6$ together with the tilted potentials (dashed dotted lines). 
		%At high driving frequencies, the probability density cannot form a plateau any more when the potential is tilted towards the 		shallow slope. In contrast, when the potential is tilted towards the steep slope, the probability density remains concentrated		around the relative minimum even at high driving frequencies. 
		The observed probability densities are in good agreement with the expected ones (shaded area). 
		At 500\,Hz, the probability densities approach the density in the non-driven case (green curves).}
	\label{fig:fig4}
\end{figure}

To model the theoretically expected jump rates and to support the observed scaling behavior, we calculated the average current $S$ at the maximum of $V(x)$ during each half period, see lines in ~\fref{fig:fig3}b). 
We note that for this choice in $x$ we expect a minimal contribution of oscillating currents that do not contribute to the drift current across periods. 
The excellent agreement with the data corroborates this assumption.
\newline
\indent
Clearly, the forward current is more severely influenced by non-linear effects. To understand this behavior it is instructive to visualize the time evolution of $S(t)$ at $f > 100$\,Hz, see \fref{fig:fig4}a) and movies S3-S5 of the SM~\cite{SM} for ratchet R1. For ratchet R2 and further frequencies see also SM10-SM12~\cite{SM}.
%The experimental data fit well to the theoretically expected profiles except for an enhanced current just after a sign change of the driving force which we attribute to a mechanical relaxation in the nanofluidic slit upon switching the capacitive force between electrodes and grounded sample. 
At {150\,Hz} the rise in the forward current is delayed by $t_d\,\approx T_f/6 = 1.1$\,ms which leads to a parabolic profile, whereas the backward current increases almost linearly. At higher frequencies the behavior is similar, simply reflecting the shorter time for the current buildup. At 250\,Hz the parabola is still visible; however, the shorter time scale cuts away most of the forward current, leading to the observed $\propto$ $f^{-2}$ current reduction. The backward current, however, still reaches almost the same value as at 150\,Hz and therefore decreases less than proportional to frequency. 
\newline
\indent
The cause for the different current buildup in forward and backward direction can be understood by again considering the evolution of $\rho(x,t)$ as shown in \fref{fig:fig4}b). 
In backward direction, an energy barrier remains in $\tilde{V}(x,t)$, which causes the accumulation of $\rho(x,t)$ and the restoration of a peak at the potential minimum. At higher frequencies a tail develops in the positive $x$ direction, and $\rho(x,t)$ approaches the static potential (green curve). In the forward direction, however, this profile has to diffuse and broaden for some time in order to reach the end of the ratchet cell at $x = L$, causing the delayed onset of the forward current. The measured delay of 1.1\,ms at 150 Hz corresponds to a drift distance of 120\,nm and a diffusion broadening at $1/e$ from 70 to 120\,nm consistent with the shape of $\rho(x)$.
The broadening also results in a quasi-linear increase of $\rho(x,t)$ between the minimum of $V(x)$ and the maximum in $\rho(x,t)$. Upon switching to the reverse direction this linear profile propagates towards the potential barrier and induces the restoration of the peak and therefore an almost immediate, linear rise of the reverse current. At higher frequencies, the maximum of $\rho(x,t)$ does not propagate so far towards the next cell in the forward direction causing a steeper linear profile of $\rho(x,t)$, and thus an even faster buildup of the reverse current (see \fref{fig:fig4}a). 
\newline
\indent
\emph{Conclusion.}---We characterized the temporal evolution of the particle probability density in a nanofluidic rocking Brownian motor for 60\,nm gold spheres with sub ms temporal and 10 nm spatial resolution.  
As predicted for such motors, we observed a current reversal at a driving frequency of 250\,Hz.
Approaching this frequency, we find that the forward current decreases $\propto f^{-2}$ whereas the backward current decreases $\propto f^{-0.5}$. This asymmetric behavior is caused by the shape of $V(x)$, which stabilizes an asymmetric probability density with a maximum close to the reverse barrier at high frequencies. Therefore, the time scale for the forward current to build up is significantly longer than for the backward current, resulting in the current reversal at intermediate time scales. 
\newline
\indent
The current reversal is governed by the finite time required for the propagation of $\rho(x,t)$ in forward direction. Thus, the transition frequency depends strongly on the applied force and only weakly on the particle potential. This behavior might open up new opportunities for the on-demand separation of nanoparticles of similar size but with a different response to the applied field (through charge or polarizability) by simply adjusting the applied frequency.
% Acknoledgements.
\begin{acknowledgments}
We would like to thank U. Drechsler for assistance in the fabrication of the glass pillars, K. M. Carrol, C. Rawlings and H. Wolf for stimulating discussions, 
and R. Allenspach and W. Riess for support. Funding was provided by the European Research Council (StG no. 307079), the European Commission FP7-ICT-2011  
no. 318804, and the Swiss National Science Foundation (SNSF no. 200020-144464 and the NCCR Molecular Systems Engineering).
\end{acknowledgments}

% References.
\bibliographystyle{apsrev4-1}

%%%%%%%%%%%% Now comes the supplemental material %%%%%%%%%%%%

% Definition and redefenition of makros.
\renewcommand{\eqref}[1]{Eq.~(\ref{#1})}
\renewcommand{\fref}[1]{Fig.~\ref{#1}}
\renewcommand{\thefigure}{S\arabic{figure}}
\renewcommand{\theequation}{S\arabic{equation}}

% Set equation counter to 0.
\setcounter{equation}{0}
\setcounter{figure}{0}

\onecolumngrid
\appendix

\section{{\large Supplemental Material}}

\section{SM1: Numerical solution of the Fokker-Planck equation}

The dynamics of an overdamped particle in a periodic potential $V(x)$ experiencing a time-dependent periodic force $F(t)$ can be described by a probability density $\rho(x,t)$ which 
obeys the Fokker-Planck equation
\begin{equation}
\partial_t \rho(x,t) = \partial_x \left[ \left( \frac{1}{\gamma}\partial_x \tilde{V}(x,t) + D_0 \partial_x \right) \rho(x,t) \right]
\label{eq:smoluchowskiSM}
\end{equation}
with $\tilde{V}(x,t) = V(x) - xF(t)$. In the following, it shall be discussed in detail how \eqref{eq:smoluchowskiSM} can be solved
numerically. To take into account the periodicity of $\rho(x,t)$, of the potential $V(x)$ and of the driving force $F(t)$,
we chose the following decomposition \cite{Denisov2009SM}
\begin{eqnarray}
\rho(x,t) &=& \sum\limits_{r,s} u_{r,s}e^{i 2\pi r x/L}e^{i s \omega t}\label{eq:rho}\\
\tilde{V}(x,t) &=& V(x) - xF(t) = \sum\limits_r b_r e^{i 2\pi r x/L} - x \sum\limits_s a_s e^{i s \omega t}
\label{eq:VSM}
\end{eqnarray}
where $L$ is the period of $V(x)$ and $\omega$ the angular frequency of $F(t)$.
The total potential $\tilde{V}(x,t) = V(x) - xF(t)$ is not periodic in space anymore but still periodic in time, as can be directly seen from 
\eqref{eq:VSM}. However, the total force $-\partial_x \tilde{V}(x,t)$ is both periodic in space and time and can therefore be written as 
\begin{equation}
\partial_x \tilde{V}(x,t) = \sum\limits_{r,s} g_{r,s}e^{i 2\pi r x/L}e^{i s \omega t}.
\label{eq:VprimeSM}
\end{equation}
To express the Fourier coefficients $g_{r,s}$ in terms of $b_r$ and $a_r$ let us consider the inverse Fourier transform 
\begin{eqnarray}
g_{r,s} &=& \frac{1}{LT} \int\limits_0^L\int\limits_0^T \left( \partial_x \tilde{V}(x,t) \right) e^{-i 2\pi r x/L}e^{-i s \omega t} dt dx
\end{eqnarray}
which can be simplified to
\begin{eqnarray}
g_{r,s} &=& \frac{1}{LT} \int\limits_0^L\int\limits_0^T  \left( \sum\limits_{r^\prime} \frac{i 2 \pi r^\prime}{L} b_{r^\prime} e^{i 2\pi {r^\prime} x/L} - 
\sum\limits_{s^\prime} a_{s^\prime} e^{i {s^\prime} \omega t} \right) e^{-i 2\pi r x/L}e^{-i s \omega t} dt dx \\
&=& \frac{1}{LT} \int\limits_0^L\int\limits_0^T  \left( \sum\limits_{r^\prime} \frac{i 2 \pi r^\prime}{L} b_{r^\prime} e^{i 2\pi ({r^\prime}-r) x/L}e^{-i s \omega t} - 
\sum\limits_{s^\prime} a_{s^\prime} e^{-i 2\pi r x/L} e^{i ({s^\prime}-s) \omega t} \right) dt dx \\
&=& \frac{1}{LT}  \sum\limits_{r^\prime} \frac{i 2 \pi r^\prime}{L} b_{r^\prime}   \int\limits_0^L  e^{i 2\pi ({r^\prime}-r) x/L} dx 
\int\limits_0^T e^{-i s \omega t} dt - \frac{1}{LT} \sum\limits_{s^\prime} a_{s^\prime} \int\limits_0^L  e^{-i 2\pi r x/L} dx 
\int\limits_0^T e^{i ({s^\prime}-s) \omega t} dt\\
&=& \frac{1}{LT}  \sum\limits_{r^\prime} \frac{i 2 \pi r^\prime}{L} b_{r^\prime} L \delta_{r,r^\prime} T \delta_{s,0} -
\frac{1}{LT} \sum\limits_{s^\prime} a_{s^\prime} L \delta_{r,0} T \delta_{s,s^\prime} \\
&=& \frac{i 2 \pi r b_r}{L} \delta_{s,0} - a_s \delta_{r,0} \label{eq:grsSM}
\end{eqnarray}
as stated in the main part of this paper. 
\newline
\indent
As first derivatives of $\rho(x,t)$ and first and second derivatives of $\tilde{V}(x,t)$ occur in \eqref{eq:smoluchowskiSM}, their Fourier decomposition shall be explicitly given:
\begin{eqnarray}
\partial_t \rho(x,t) &=&  i  \omega \sum\limits_{r,s} s u_{r,s}e^{i 2\pi r x/L}e^{i s \omega t} \label{eq:derivatives_1SM}\\
\partial_x\rho(x,t) &=&  \frac{i 2 \pi}{L} \sum\limits_{r,s} r u_{r,s}e^{i 2\pi r x/L}e^{i s \omega t} \label{eq:derivatives_2SM}\\
\partial^2_x \rho(x,t) &=&  \frac{-4 \pi^2}{L^2} \sum\limits_{r,s} r^2 u_{r,s}e^{i 2\pi r x/L}e^{i s \omega t} \label{eq:derivatives_3SM}\\
\partial^2_x \tilde{V}(x,t) &=& \frac{i 2 \pi}{L} \sum\limits_{r,s} r g_{r,s} e^{i 2\pi r x/L}e^{i s \omega t}.\label{eq:derivatives_4SM}
\end{eqnarray}
Insertion of \eqref{eq:derivatives_1SM}-(\ref{eq:derivatives_4SM}) into the Fokker-Planck equation \eqref{eq:smoluchowskiSM} delivers
\begin{eqnarray}
i \omega \sum\limits_{r,s} s u_{r,s}e^{i 2\pi r x/L}e^{i s \omega t} &=& \frac{i 2 \pi}{\gamma L} \sum\limits_{m,n} m g_{m,n} e^{i 2\pi m x/L}e^{i n \omega t} 
\sum\limits_{r,s} u_{r,s}e^{i 2\pi r x/L}e^{i s \omega t} + \nonumber\\
&+& \frac{i 2 \pi}{\gamma L} \sum\limits_{m,n} g_{m,n}e^{i 2\pi m x/L}e^{i n \omega t} \sum\limits_{r,s} r u_{r,s}e^{i 2\pi r x/L}e^{i s \omega t} - 
\frac{4 \pi^2 D_0}{ L^2} \sum\limits_{r,s} r^2 u_{r,s}e^{i 2\pi r x/L}e^{i s \omega t}\\
&=&  \frac{i 2 \pi}{\gamma L} \sum\limits_{r,s} \sum\limits_{m,n} m g_{m,n} u_{r,s} e^{i 2\pi (r+m) x/L}e^{i (s+n) \omega t} + \nonumber\\
&+&  \frac{i 2 \pi}{\gamma L} \sum\limits_{r,s} \sum\limits_{m,n} r g_{m,n} u_{r,s} e^{i 2\pi (r+m) x/L}e^{i (s+n) \omega t} - 
\frac{4 \pi^2 D_0} {L^2} \sum\limits_{r,s} r^2 u_{r,s}e^{i 2\pi r x/L}e^{i s \omega t}\\
&=&  \frac{i 2 \pi}{\gamma L} \sum\limits_{r,s} \sum\limits_{m,n} m g_{m,n} u_{r-m,s-n} e^{i 2\pi r x/L}e^{i s \omega t} + \nonumber\\
&+&  \frac{i 2 \pi}{\gamma L} \sum\limits_{r,s} \sum\limits_{m,n} (r-m) g_{m,n} u_{r-m,s-n} e^{i 2\pi r x/L}e^{i s \omega t} -
\frac{4 \pi^2 D_0}{ L^2} \sum\limits_{r,s} r^2 u_{r,s}e^{i 2\pi r x/L}e^{i s \omega t}\\
&=&  \frac{i 2 \pi}{\gamma L} \sum\limits_{r,s} \sum\limits_{m,n} r g_{m,n} u_{r-m,s-n} e^{i 2\pi r x/L}e^{i s \omega t} - 
\frac{4 \pi^2 D_0}{ L^2} \sum\limits_{r,s} r^2 u_{r,s}e^{i 2\pi r x/L}e^{i s \omega t}
\end{eqnarray}
which can be transformed into
\begin{equation}
\sum\limits_{r,s} \left( i \gamma \omega s u_{r,s}  + \frac{4 \pi^2 D_0 \gamma}{ L^2} r^2 u_{r,s}  - \frac{i 2 \pi}{L} \sum\limits_{m,n} r g_{m,n} u_{r-m,s-n} \right) e^{i 2\pi r x/L}e^{i s \omega t} = 0.
\label{eq:almostdoneSM}
\end{equation}
As \eqref{eq:almostdoneSM} has to be fulfilled $\forall t$ and $\forall x \in [0,L]$ the following recurrence relation is obtained
\begin{equation}
\left( i \gamma \omega s  + \frac{4 \pi^2 D_0 \gamma}{ L^2} r^2 \right) u_{r,s}  - \frac{i 2 \pi}{L} \sum\limits_{m,n} r g_{m,n} u_{r-m,s-n} = 0.
\label{eq:recrelationSM}
\end{equation}
Using \eqref{eq:grsSM} delivers the final result
\begin{equation}
\left( i \gamma \omega s  + \frac{4 \pi^2 D_0 \gamma}{ L^2} r^2 \right) u_{r,s}  - \frac{i 2 \pi}{L}  \sum\limits_{m,n} r  \left( \frac{i 2 \pi m b_m}{L} \delta_{n,0} - a_n \delta_{m,0} \right) u_{r-m,s-n} = 0.
\label{eq:finalresultSM}
\end{equation}
To determine the unknown coefficients $u_{r,s}$, \eqref{eq:finalresultSM} is transformed into a matrix equation by mapping 
the double index variable $u_{r,s}$ onto a single index variable $\tilde{u}_l$ according to $(r,s) \rightarrow 1 + (r+R)(2S+1) + s + S$.
Therefore, two cutoff parameters $S$ and $R$ for the spatial and temporal Fourier coefficients have to be chosen to restrict the range
of the indices $s$ and $r$ to $-S < s < S$ and $-R < r < R$. 
By using the normalization condition of the probability density $\int_0^L \rho(x,t) dx = 1  \,\,\, \forall t$ and by noting that  
$u_{r,s} =  u_{-r,-s}$ since $\rho(x,t) \in \mathbb{R}$, it follows that $ u_{0,0} = L^{-1}$ and we can write the
obtained matrix equation into the form  $Ay = b$. The unique solution can then be easily found by using standard linear algebra routines
like e.g. numpy.linalg.solve from the numpy package for Python.

We find, the quality of the numerical solution of the Fokker-Planck equation depends critically on choosing sufficiently large cutoff parameters. 
In all our calculations we used $S=60$ and $R \geq 60$  which turned out to deliver satisfactory results as a further increase of $S$ or $R$
did not alter the solution.

\section{SM2: Nanoparticles}

We used high optical density citrate stabilized spherical gold nanoparticles of $60$\,nm size purchased from BBI solutions 
(product code  HD.GC60.OD100) for our experiments. According to the manufacturer, the coefficient of variation was~$8$\,\% and 
the particle concentration was $2.6 \times 10^{12}$ per ml. As the concentration was too high for our experiments,
we diluted the dispersion by a factor of 15 with ultrapure water (Millipore, $18$\,M$\Omega$cm).

\section{SM3: Sample preparation}

For all experiments described in this paper, we used highly doped silicon wafers which were first coated with a layer of HM8006
(JSR Inc.) to increase the adhesion of polypthalaldehyde (PPA) in water. The pre-formulated solution of HM8006
was spincoated at 6000\,rpm for 35\,s and then the samples were baked for 90\,s on a hotplate at $225^{\circ}$ for cross-linking. 
The resulting layer had a thickness of approximately $55$\,nm as measured by an AFM across a scratch.
In the next step, the sample was coated with the thermally sensitive polymer polypthalaldehyde (PPA). The material was purchased from
the IBM Reaserch lab at Almaden, USA where it was synthesized by J.~Hedrick and co-workers~\cite{Coulembier2009SM}. The PPA was 
spincoated at 4250\,rpm for 35\,s with subsequent curing at $90^{\circ}$ for 3\,min on a hotplate to evaporate residual solvent. 
As before, a thickness of $157$\,nm of the PPA layer was measured by AFM across a scratch.

\section{SM4: Stability of the experimental apparatus}

\begin{figure}[ht]
	\centering
	\includegraphics[width=0.34\textwidth]{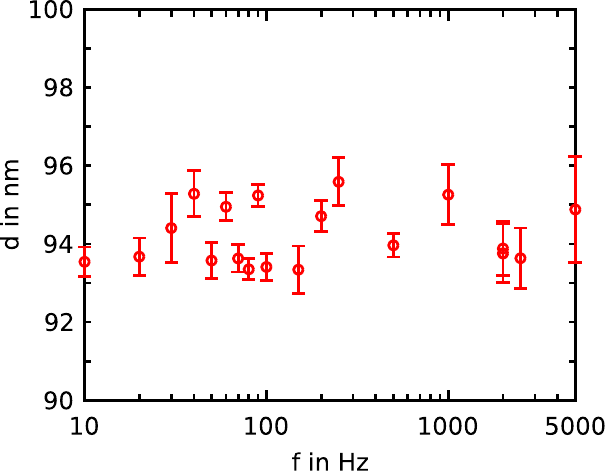}
	\caption{The high stability of our experimental apparatus allowed us to perform a
		multitude of experiments under practically the same conditions. The overall drift of the gapdistance was of the
		order of 2\,nm corresponding to a variation of 2\,\%.}
	\label{fig:fig1_som}
\end{figure}

\FloatBarrier
%\clearpage
\section{SM5: Details on particle numbers}

\begin{table}[h!]
	\centering		
	\begin{tabular}{l|c|c|c|c|c|c|c|c|c|c|c|c|c|c}
		&  \,\,10\,Hz\,\, & \,\,20\,Hz\,\, & \,\,30\,Hz\,\, & \,\,40\,Hz\,\, & \,\,50\,Hz\,\, & \,\,60\,Hz\,\, & \,\,70\,Hz\,\, & \,\,80\,Hz\,\, & \,\,90\,Hz\,\, & \,\,100\,Hz\,\, & \,\,150\,Hz\,\, & \,\,200\,Hz\,\, & \,\,250\,Hz\,\, & \,\,500\,Hz\,\, \\ \hline\hline
		D1 & 2 & 2 & 2 & 2 & 2 & 2 & 2 & 2 & 2 & 2 & 2 & 2 & 1 & 1\\ \hline
		R1/R2 & 5 & 5 & 5 & 5 & 5 & 5 & 5 & 5 & 5 & 5 & 5 & 5 & 5 & 5  \\ \hline
		D2 & 2 & 2 & 2 & 2 & 2 & 2 & 2 & 3 & 3 & 3 & 3 & 3 & 4 & 4 \\ \hline
	\end{tabular}
	\caption{Number of gold nanospheres in the respective structures of the patterned ratchets (see also Fig. 1b) of the main part) for driving frequencies between 10 and 500\,Hz.}
	\label{my-label}
\end{table}

\FloatBarrier
\clearpage
\section{SM6: Determination of the experimental potentials}

\begin{figure}[ht]
	\centering
	\includegraphics[width=0.55\textwidth]{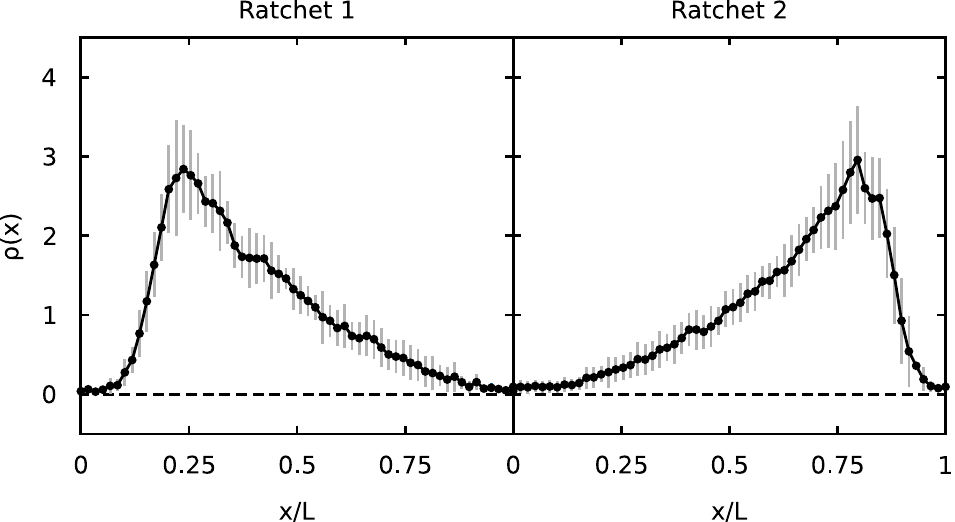}
	\caption{Experimentally observed average probability density for ratchets R1 and R2 in the static case. The error
		bars represent the standard deviation of the normalized probability densities of the single teeth.}
	\label{fig:fig2_som}
\end{figure}

To simulate the theoretically expected propagation of the probability density $\rho(x,t)$, it was necessary to know the ratchet potential. 
Therefore, in a first step, the probability density in the non-driven case was 
determined for both ratchets, see \fref{fig:fig2_som}.

\begin{figure}[ht]
	\centering
	\includegraphics[width=0.45\textwidth]{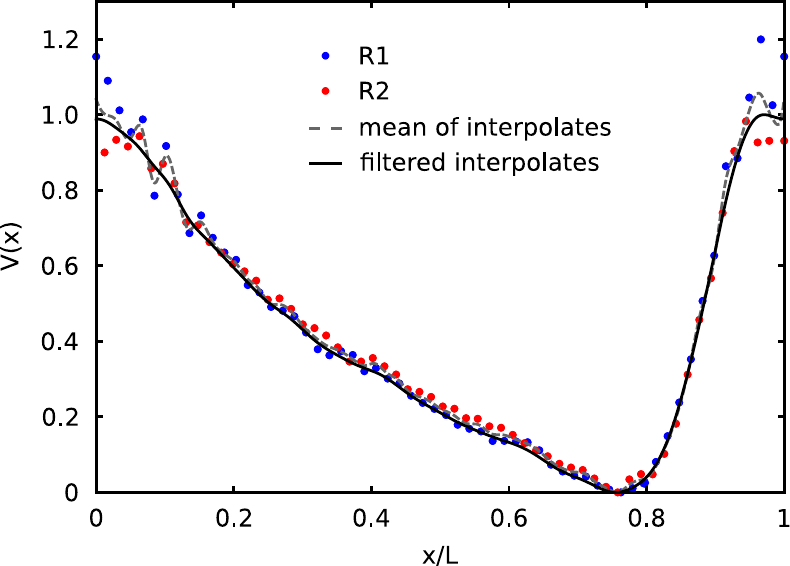}
	\caption{Interpolated potentials of ratchet R1 (blue) and R2 (red), their average (dashed line)
		and the average after applying a Gaussian filter to remove noise. The scaling of the potentials is such that $\Delta V = 1$ of the filtered
		potential.}
	\label{fig:fig3_som}
\end{figure}

As discussed in the main part, we used Boltzmann's principle to infer the ratchet potentials of R1 and R2 from the
observed probability densities.
Evidently, the measured ratchet potentials of R1 and R2 are very similar (see Fig. 1c) of the main part) and we therefore used an average 
potential for all our calculations. To determine the average, we first interpolated
the experimental potentials with a resolution of 1\,nm. Then, we mirrored the potential of ratchet R2. The best
overlap of the two potentials was achieved for a relative shift of 13\,nm between the ratchets as determined via cross 
correlation. To remove noise, a Gaussian filter with a standard deviation of its kernel of 11\,nm was applied to the average, 
see \fref{fig:fig3_som}. After rescaling, the smoothed average potential was used for all simulations throughout the paper.

\FloatBarrier
\section{SM7: Details on jump rate analysis in ratchets R1 and R2}

\begin{figure}[ht]
	\centering
	\includegraphics[width=0.80\textwidth]{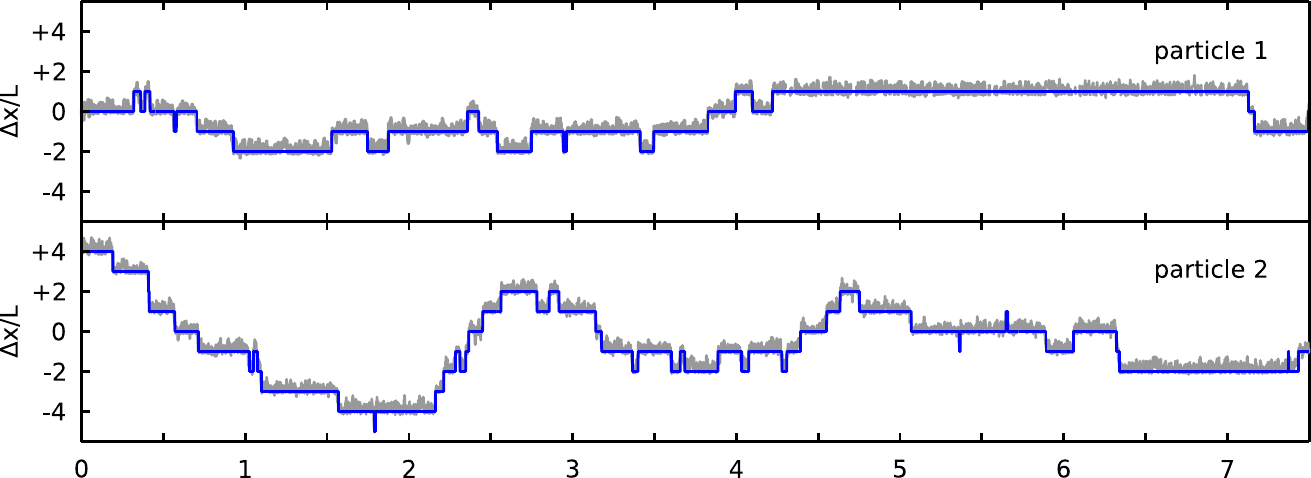}
	\caption{Time traces of two particles located in ratchet R1 without external driving (gray curves) together with
		the position of the potential minimum of the respective ratchet tooth (blue curves). As expected for potential barriers $\gg 1\,k_BT$, 
		only jumps of a single tooth length can occur. From the 77 jumps that were detected, a jump rate $r = 5.5 \pm 0.7$\,Hz was calculated.
		Note that only trajectories spanning over at least 20 consecutive frames were considered.}
	\label{fig:fig4_som}
\end{figure}

In the main part of the paper, the jump rate of the particles between the ratchet teeth was used to cross-check the tooth height as inferred 
from Boltzmann's principle. Here, a more detailed description of how the jump rate was determined shall be given. First, a 7.5\,s long video 
of the particles trapped in ratchet R1 and R2 without external driving at a frame rate of 4000\,fps was recorded. Then, particle detection 
and tracking was performed and the measured 2D particle trajectories were collapsed on the x-axis. The resulting time traces for two particles 
in ratchet R1 are displayed in \fref{fig:fig4_som}. Both traces show a clearly visible step-like shape with the steps indicating
jumps. In total, 77 jumps were observed which corresponds to a jump rate $r = 5.5 \pm 0.7$\,Hz assuming a Poissonian error for the number 
of detected jumps.

\begin{figure}[ht]
	\centering
	\includegraphics[width=0.60\textwidth]{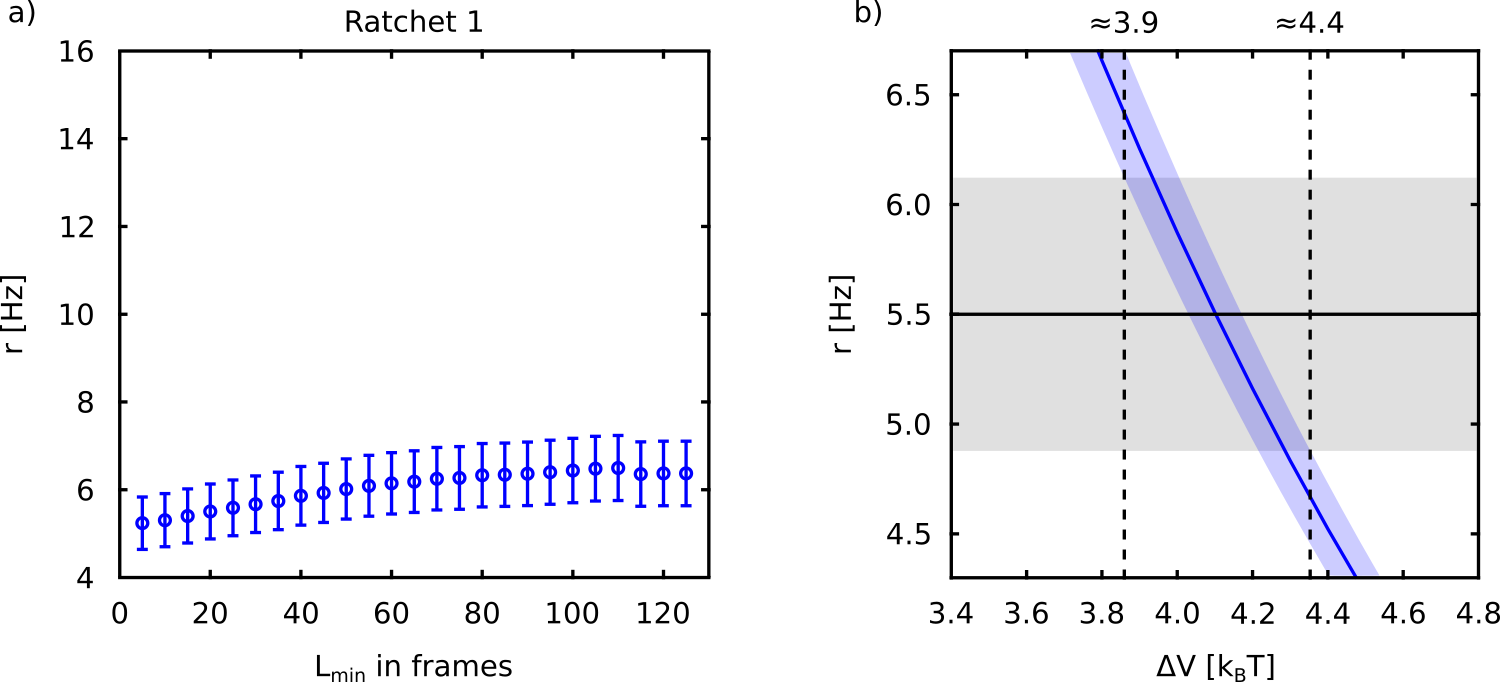}
	\caption{a) Calculated jump rate in ratchet R1 for different values of the minimal trajectory length $L_{\rm min}$. 
		b) Expected jump rates for the experimentally measured diffusion constant of $4.2\,\mu$m$^2/$s in dependence
		of the tooth height $\Delta V$ (blue curve) in comparison with the measured rate of $5.5$\,Hz (black curve). The intersection point of the
		blue an black curve determines the tooth height which is in best agreement with the experimental jump rate. If the measurement
		inaccuracies of the jump rate and the diffusion constant are taken into account (shaded areas), a tooth height between $3.9\,k_BT$
		and $4.4\,k_BT$ is most plausible.}
	\label{fig:fig5_som}
\end{figure}

\begin{figure}[ht]
	\centering
	\includegraphics[width=0.60\textwidth]{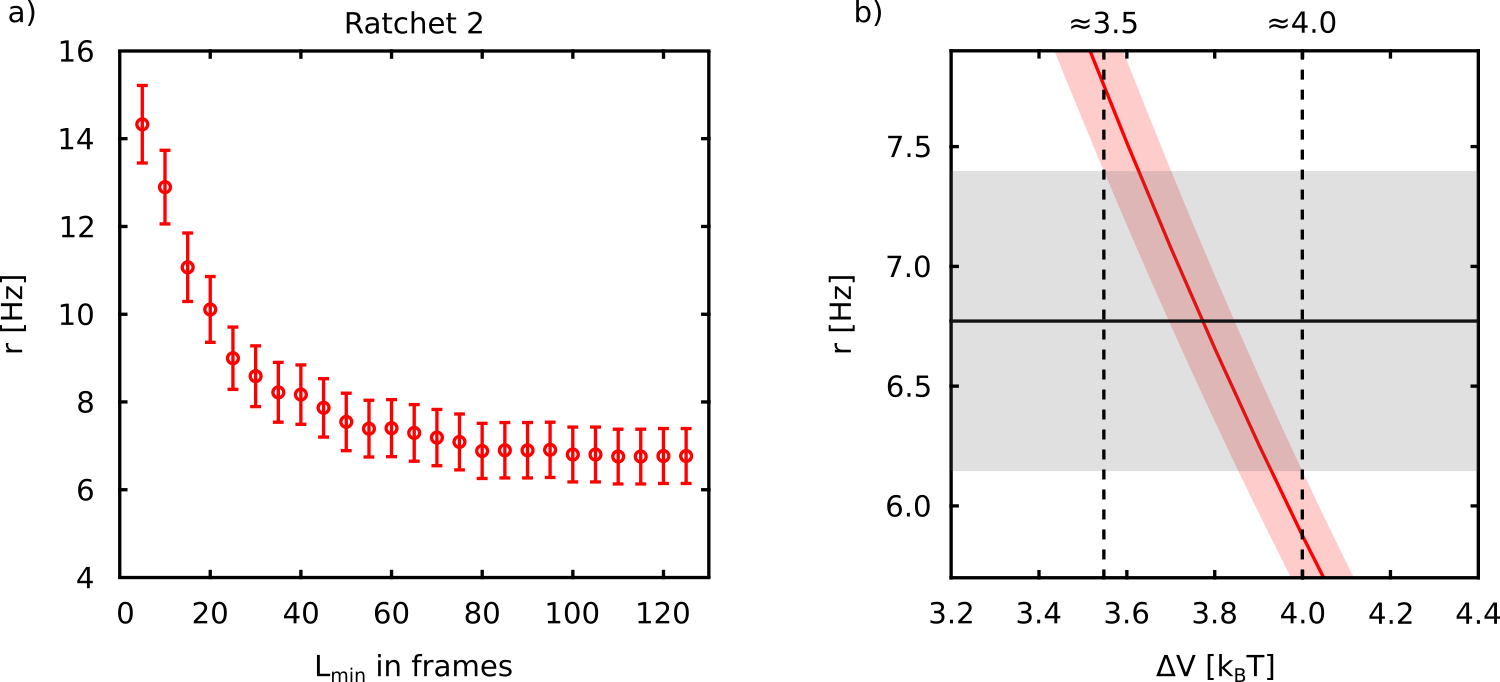}
	\caption{a) Calculated jump rate $r$ in ratchet R2 for different values of the minimal trajectory length $L_{\rm min}$. 
		Obviously, $r$ shows a strong dependence on $L_{\rm min}$ and stabilizes at a value of $r \approx 6.8$\,Hz for $L_{\rm min} \geq 80$.   
		b) Expected jump rates for the experimentally measured diffusion constant of $4.2\,\mu$m$^2/$s in dependence
		of the tooth height $\Delta V$ (blue curve) in comparison with a rate of $6.8$\,Hz (black curve). The intersection point of the
		blue an red curve determines the tooth height which is in best agreement with the experimental jump rate. If the measurement
		inaccuracies of the jump rate and the diffusion constant are taken into account (shaded areas), a tooth height between $3.5\,k_BT$
		and $4.0\,k_BT$ is most plausible.}
	\label{fig:fig6_som}
\end{figure}

As already stated in the main part, the jump rate $r$ is proportional to the effective diffusion constant $r = 2 D_{eff}/L^2$~\cite{Ferrando1992SM}
with $D_{eff} = D_0 / \left[ \int_0^L \exp\left(\frac{V(x)}{k_BT}\right) \int_0^L \exp\left(-\frac{V(x)}{k_BT}\right) \right]$ 
where $D_0$ is the free diffusion constant~\cite{Lifson1962SM}.
By combining the two expressions, the jump rate can be expressed as
\begin{equation}
r = \frac{2 D_0}{L^2 \int\limits_0^L \exp\left(\frac{V(x)}{k_BT}\right) \int\limits_0^L \exp\left(-\frac{V(x)}{k_BT}\right)}.
\label{eq:jumprate}
\end{equation}
Then, the expected jump rate for different tooth heights $\Delta V$ can be calculated by linear scaling of the potential.
\newline
\indent
At this point, a short comment on the experimental determination of the jump rate $r$ is advisable. As discussed above, 
the most straightforward way to determine $r$ is to measure the particle trajectories and then count the number of jumps. However, this ansatz is only 
meaningful if the trajectories of the particles can be measured with good accuracy and if  $r$ does not depend on the minimal chosen 
trajectory length $L_{\rm min}$. To check if this is really the case, $r$ was determined for ratchet R1 for different values of  $L_{\rm min}$, see \fref{fig:fig3_som}a). 
As required, $r$ is practically independent on $L_{\rm min}$ and therefore, $\Delta V$ can be determined 
from \eqref{eq:jumprate} and $r = 5.5 \pm 0.7$\,Hz.
As one can directly see in \fref{fig:fig5_som}b), the tooth height of ratchet R1 lies most likely between 
$4.1\,k_BT$ and $4.6\,k_BT$ when the experimental errors of measuring $D_0$ and $r$ are taken into account.
\newline
\indent
In the main part of the paper, it was also argued that for ratchet R2 the tooth height $\Delta V$ could not 
be determined from the jump rate because two particles were trapped in neighboring teeth. The problem
that is caused by this is the following. As our optical setup is diffraction limited, the observed size 
of a 60\,nm gold particle is around 200\,nm. Therefore, if two particles come too close together, only a single 
image is observed which impedes single particle detection. Hence, it is not possible to
distinguish between two particles that pass by each other and two particles that are repelled from
each other. When the measured particle positions are then linked to trajectories it is not clear which
of the two cases is chosen by the linking algorithm and often the particle positions cannot be linked to
trajectories at all. This fact is reflected by a large number of very short trajectories and by a strong
dependence of the calculated jump rate $r$ on the the minimal trajectory length $L_{\rm min}$, see 
\fref{fig:fig6_som}a). Only for trajectory lengths $L_{\rm min} \geq 80$, the jump rate approaches a static
value of $r \approx 6.8$\,Hz, see \fref{fig:fig6_som}b). 
However, since the origin of the decreasing rate is unclear, we decided not to use the rate of ratchet R2 to determine the tooth height $\Delta V$.

%Finally, it has to be noted that the interaction between particles trapped in neighboring teeth can also 
%modify the jump rate. Unfortunately, as argued above, we are unsure about the reliability of the particle
%trajectories and cannot make a statement if the observed behavior in \fref{fig:fig6_som}a) is already a manifestation
%of particle particle interaction.

\section{SM8: Diffusion in narrow channels}

In our experiments, the gap distance was approximately 100\,nm, the ratchet was embedded by 30\,nm and the tooth height was 30\,nm (see Fig. 1b) and the main text). Hence, at the top of a tooth, the distance to the cover slip was 130\,nm and at its bottom the distance was 160\,nm. 
It is known that in corrugated channels, the diffusivity is position dependent~\cite{Yang2017}. Furthermore, we measured the diffusion of 60\,nm gold spheres in the nanofluidic slit \cite{Fringes2018} as a function of the gap distance, albeit at a higher Debye length. From this data we expect that the diffusivity at the top of a tooth is $\approx 10\,\%$ lower than in the centre of the ratchet (at $145\,$nm gap distance).
\newline
\indent
For the observed rates this means that we expect lower experimental counts than expected from the mean diffusion constant. Therefore, we will slightly underestimate the potential using Kramer's method.
\newline
\indent
For the effect on the particle drift velocity, however, it must be noted that not only the diffusivity is lower at the top of a tooth but that the force experienced by a particle is also larger by $\approx 10\,\%$, due to the slightly increased speed of the laminar flow (volume conservation). The lower diffusivity at the top impedes a jump to a neighboring tooth whereas the higher force facilitates it.
As these two effects counteract each other, their overall effect on particle dynamics can be neglected to first order.
The excellent agreement between our experimental results and the theoretical model corroborates this interpretation.

\FloatBarrier
\clearpage
\section{SM9: Time resolved analysis of ratchet R2}

\begin{figure}[ht]
	\centering
	\includegraphics[width=0.48\textwidth]{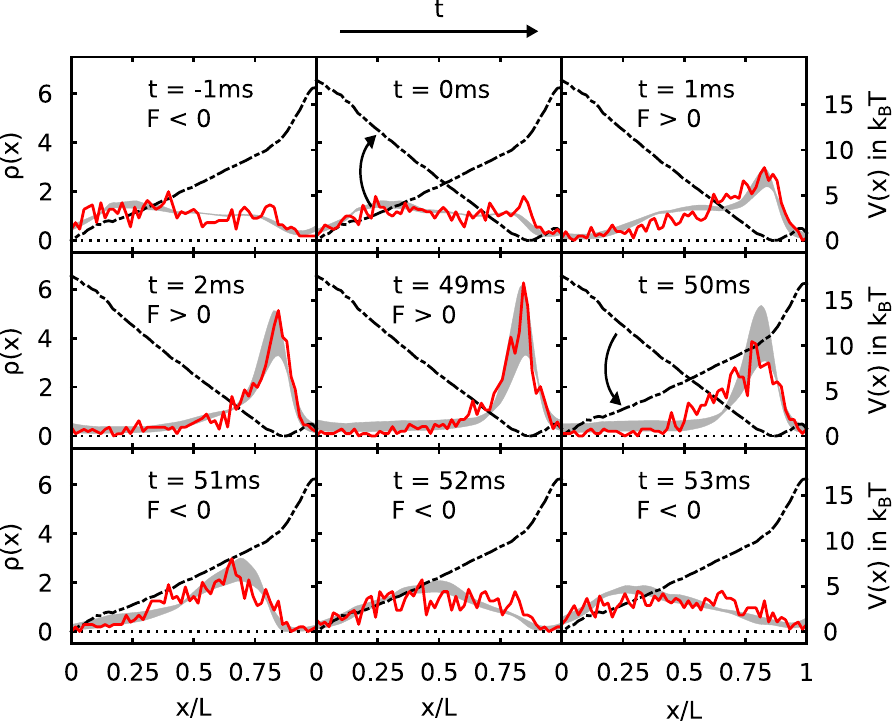}
	\caption{Time evolution of the probability density (red) inside ratchet R2 for square wave driving at 10\,Hz and 4\,V. 
		At $t=0$\,ms, the force changes from negative to positive and at $t=50$\,ms it changes back to negative again. The two
		configurations correspond to a forward and backward tilt of the potential (dashed dotted line). After each sign change, 
		the motor needs $\approx$\,$4$\,ms to follow the change in force. The gray shaded areas show the theoretically expected evolution 
		taking into account the inaccuracies in determining the tooth height, the diffusivity and the rocking force.}
	\label{fig:fig7_som}
\end{figure}

\section{SM10: Current buildup in ratchet R2}

\begin{figure}[ht]
	\centering
	\includegraphics[width=0.45\textwidth]{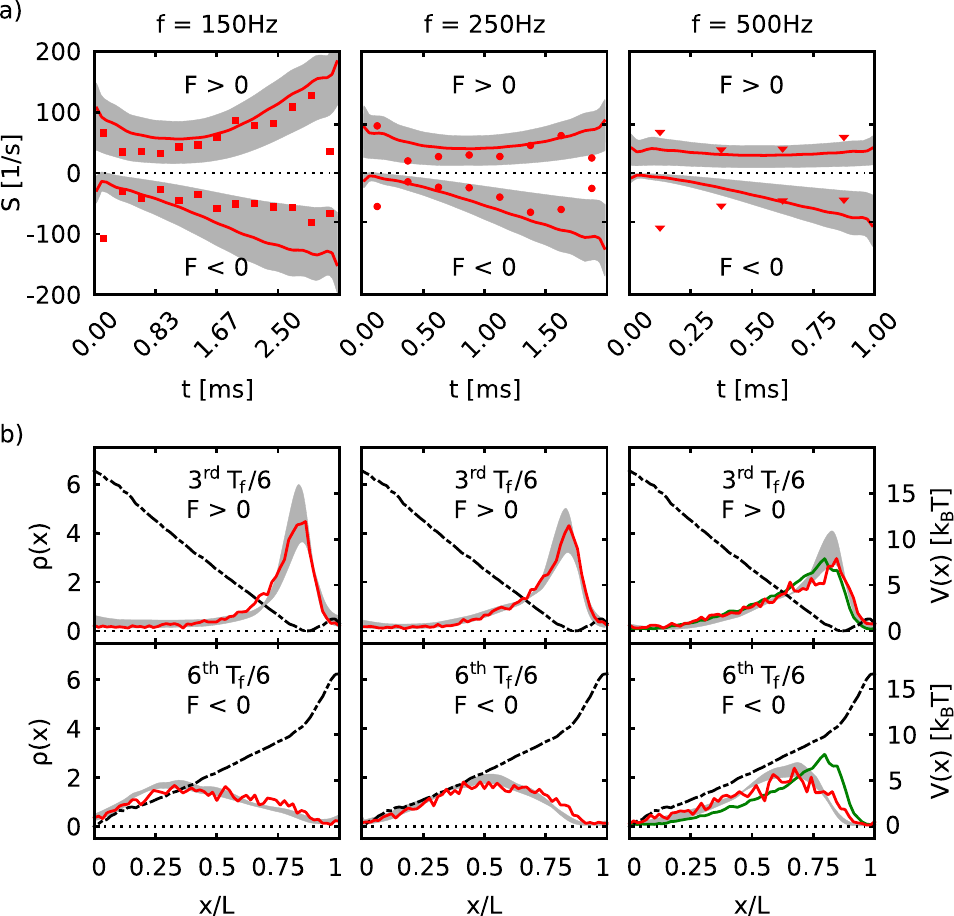}
	\caption{a) Buildup of the current in ratchet R2 at the maximum of $V(x)$ for frequencies of 150\,Hz (square), 
		250\,Hz (circles) and 500\,Hz (triangles) in comparison with our theoretical model (red lines) taking into account
		experimental inaccuracies (gray shaded area). 
		b) Probability densities averaged over $T_f/6$ together with the tilted potentials (dashed dotted lines). 
		The observed probability densities are in good agreement with the expected ones (shaded area). 
		At 500\,Hz, the probability densities approach the density in the non-driven case (green curves).}
	\label{fig:fig8_som}
\end{figure}

\FloatBarrier
\clearpage
\section{SM11: Additional measurements on ratchet R1}

\begin{figure}[ht]
	\centering
	\includegraphics[width=0.95\textwidth]{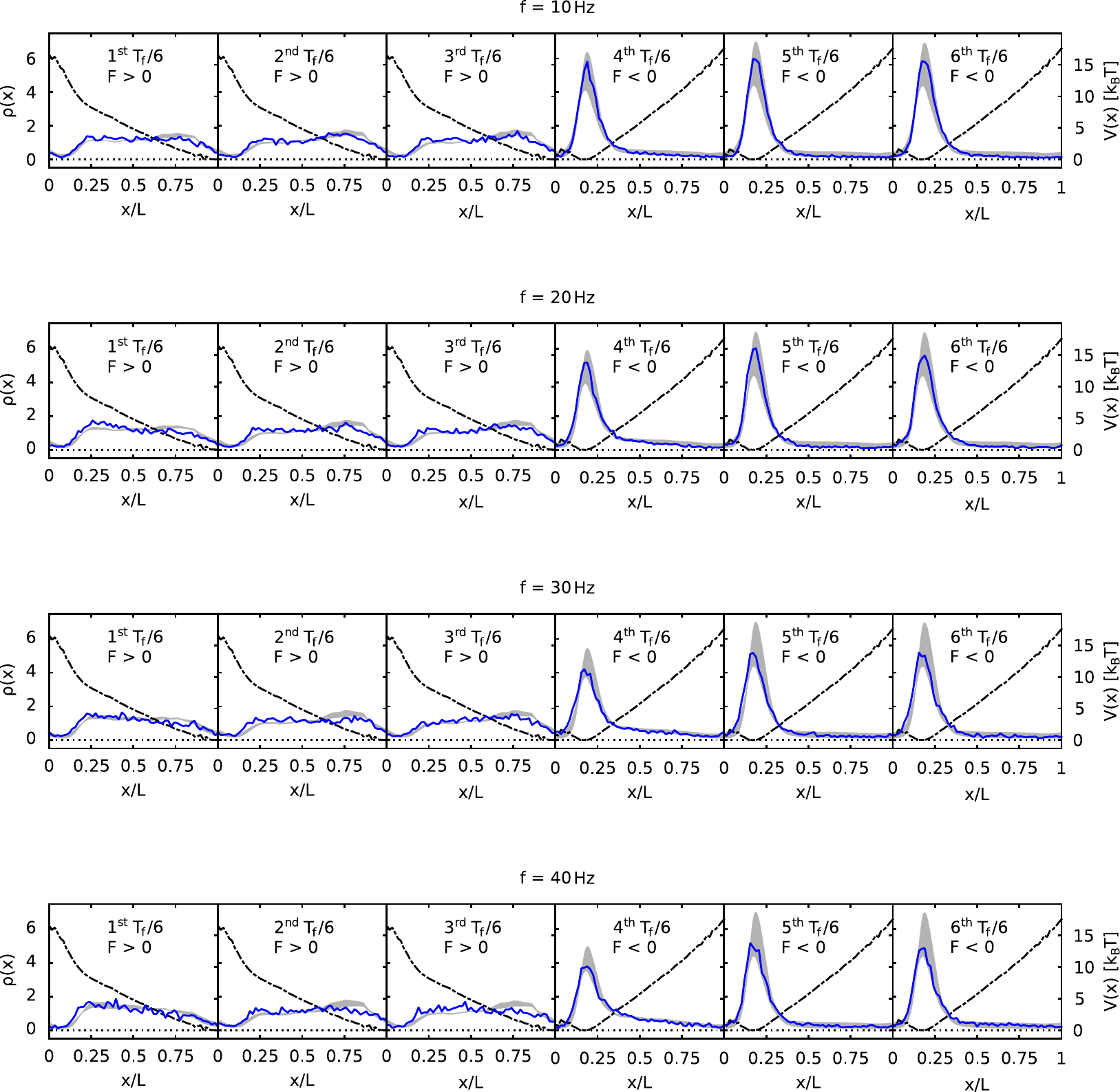}	
	\caption{Measured propagation of the probability density (blue) averaged over intervals of $T_f/6$ for driving frequencies from $f=10$\,Hz to $f=40$\,Hz 
		in comparison to the theoretically expected propagation taking into account experimental inaccuracies (gray shaded area).}
	\label{fig:fig9_som}
\end{figure}
\begin{figure}[ht]
	\centering
	\includegraphics[width=0.95\textwidth]{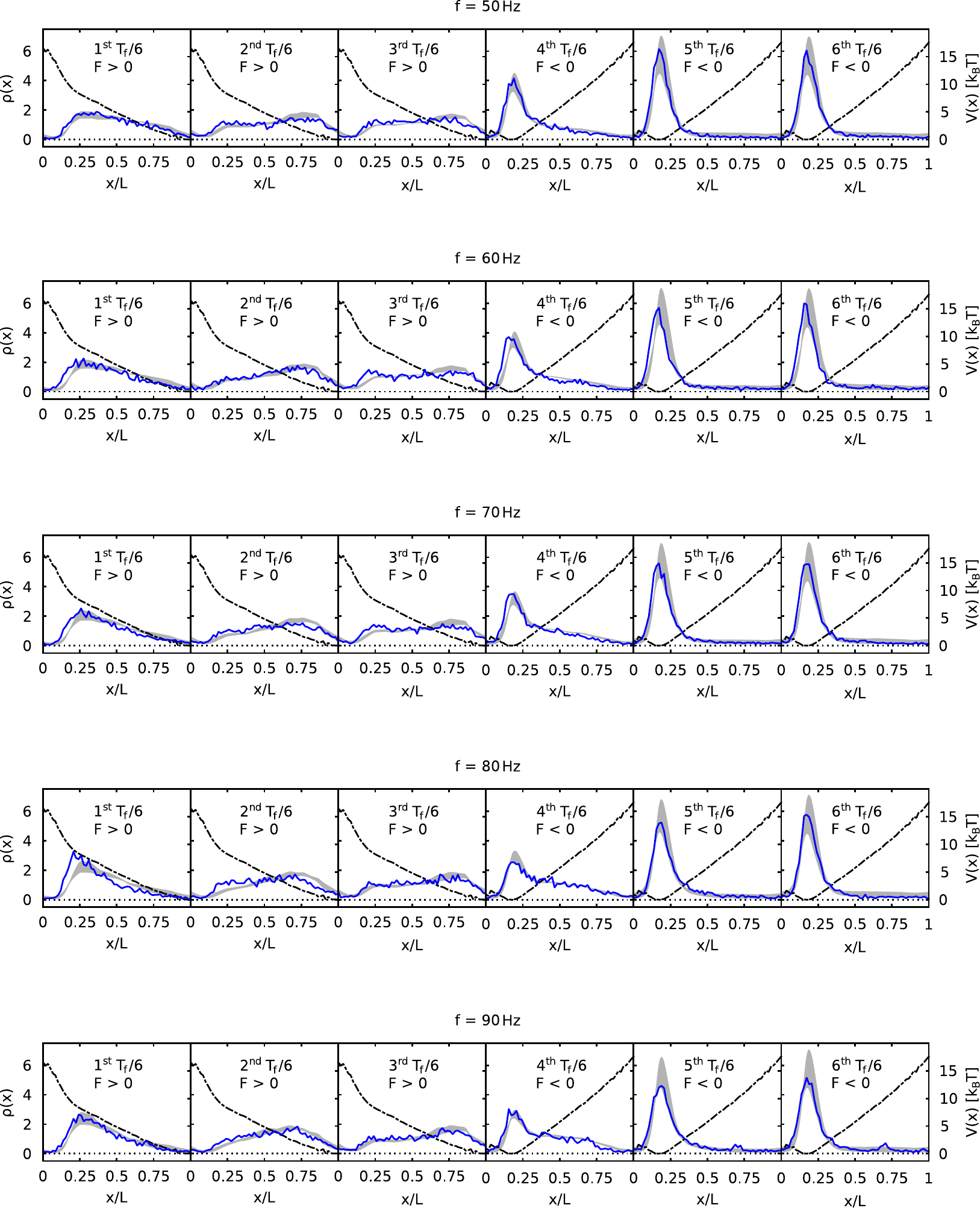}
	\caption{Measured propagation of the probability density (blue) averaged over intervals of $T_f/6$ for driving frequencies from $f=50$\,Hz to $f=90$\,Hz 
		in comparison to the theoretically expected propagation taking into account experimental inaccuracies (gray shaded area).}
	\label{fig:fig10_som}
\end{figure}
\begin{figure}[ht]
	\centering
	\includegraphics[width=0.95\textwidth]{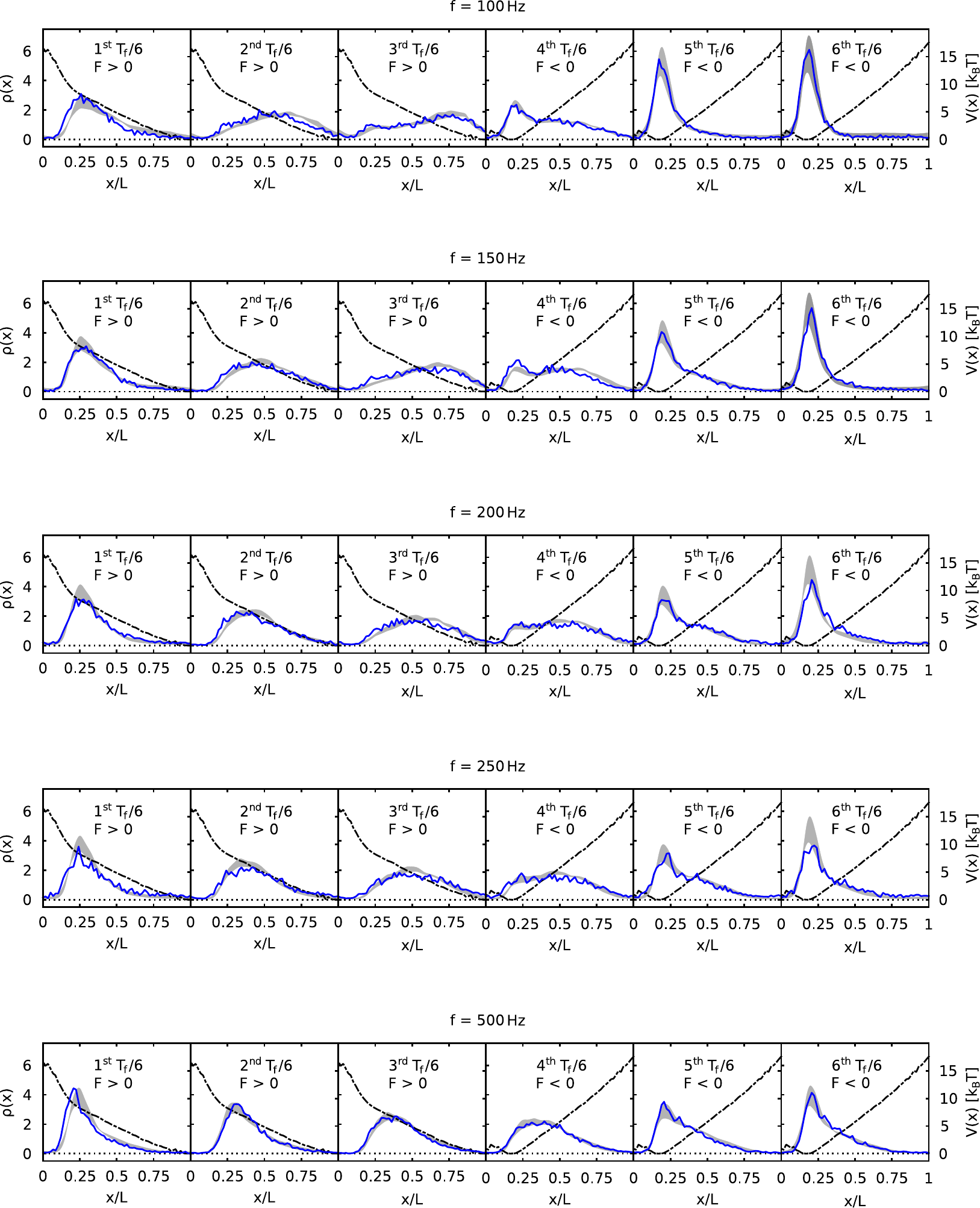}
	\caption{Measured propagation of the probability density (blue) averaged over intervals of $T_f/6$ for driving frequencies from $f=100$\,Hz to $f=500$\,Hz 
		in comparison to the theoretically expected propagation taking into account experimental inaccuracies (gray shaded area).}
	\label{fig:fig11_som}
\end{figure}

\FloatBarrier
\clearpage
\section{SM12: Additional measurements on ratchet R2}

\begin{figure}[ht]
	\centering
	\includegraphics[width=0.95\textwidth]{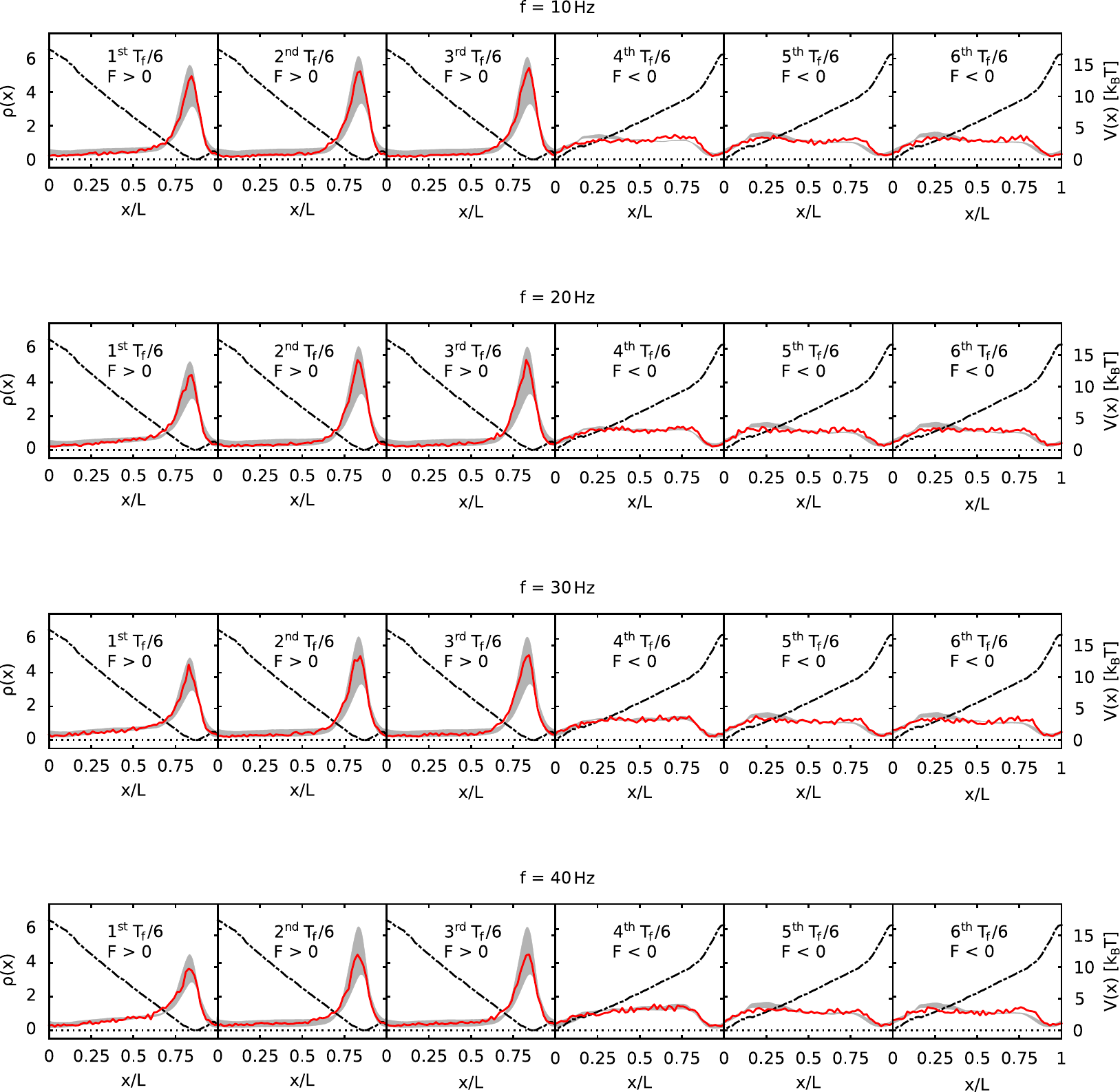}
	\caption{Measured propagation of the probability density (red) averaged over intervals of $T_f/6$ for driving frequencies from $f=10$\,Hz to $f=40$\,Hz 
		in comparison to the theoretically expected propagation taking into account experimental inaccuracies (gray shaded area).}
	\label{fig:fig12_som}
\end{figure}
\begin{figure}[ht]
	\centering
	\includegraphics[width=0.95\textwidth]{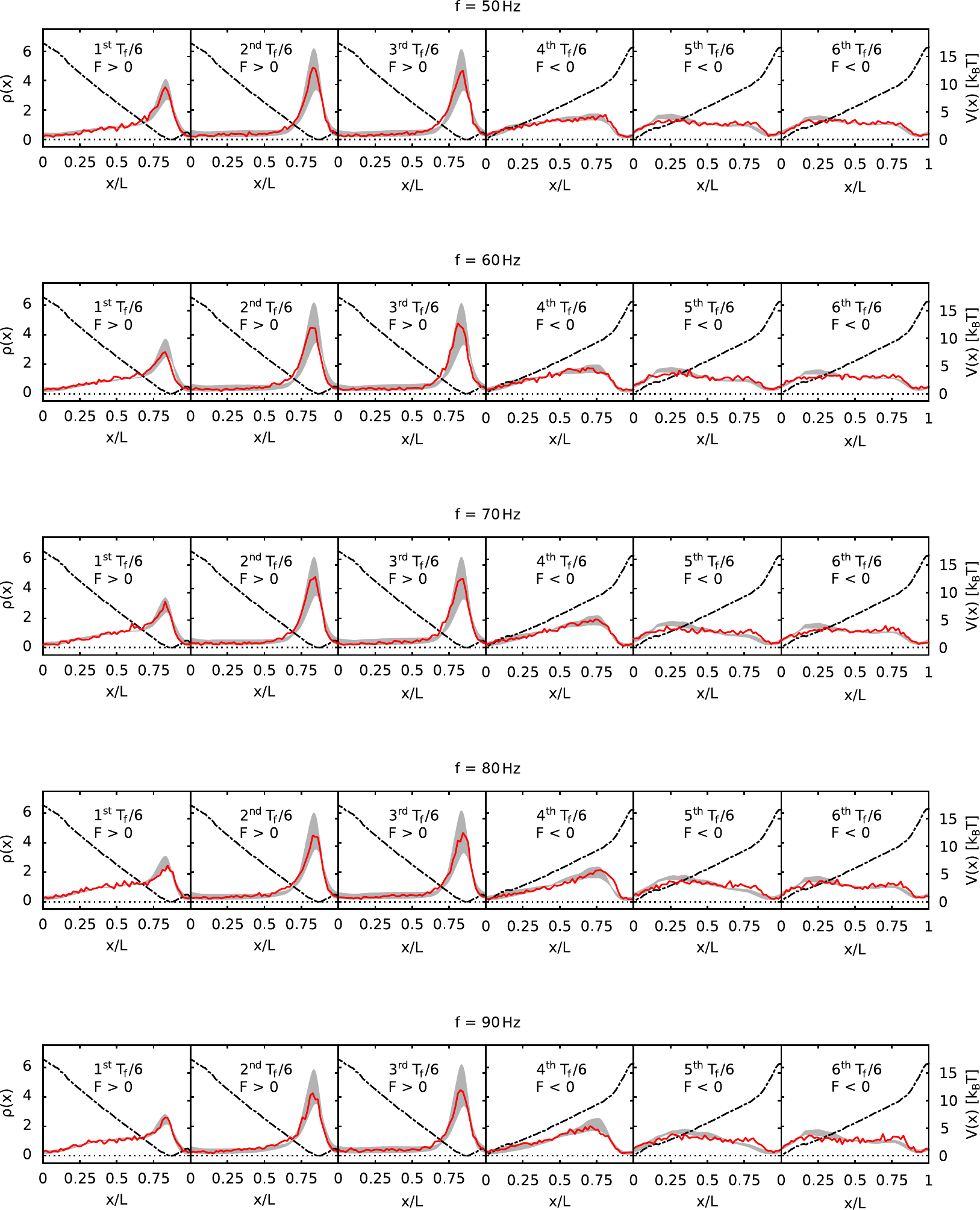}
	\caption{Measured propagation of the probability density (red) averaged over intervals of $T_f/6$ for driving frequencies from $f=50$\,Hz to $f=90$\,Hz 
		in comparison to the theoretically expected propagation taking into account experimental inaccuracies (gray shaded area).}
	\label{fig:fig13_som}
\end{figure}
\begin{figure}[ht]
	\centering
	\includegraphics[width=0.95\textwidth]{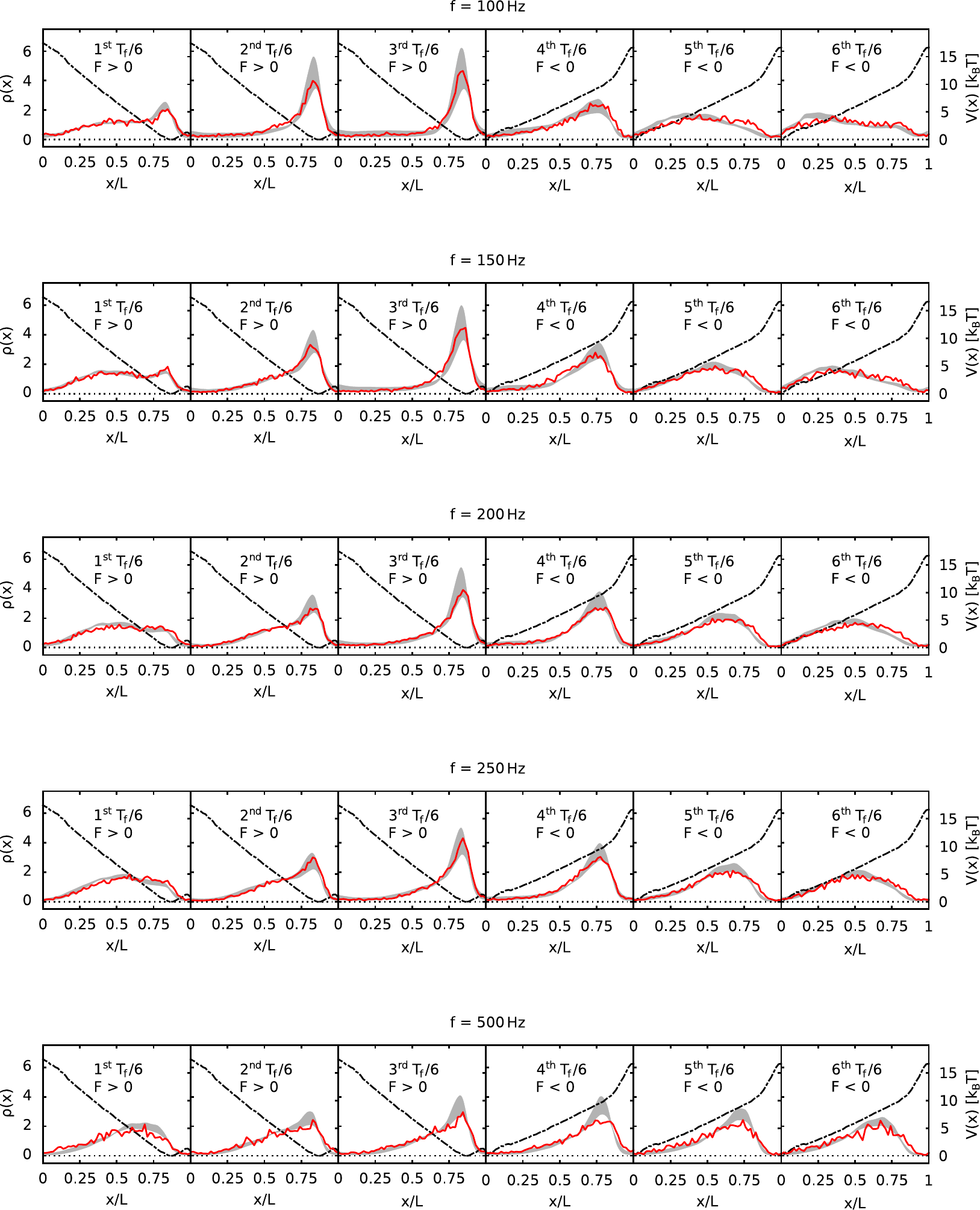}
	\caption{Measured propagation of the probability density (red) averaged over intervals of $T_f/6$ for driving frequencies from $f=100$\,Hz to $f=500$\,Hz 
		in comparison to the theoretically expected propagation taking into account experimental inaccuracies (gray shaded area).}
	\label{fig:fig14_som}
\end{figure}

\FloatBarrier

\end{document}